\documentclass[aps,prb,superscriptaddress,twocolumn,tightenlines,showpacs,floatfix,amsmath,amssymb]{revtex4-1}
\usepackage{graphicx}
\usepackage{physics}
\usepackage{dsfont}
\usepackage{hyperref}

\begin{document}

\title{Environment-assisted invariance does not necessitate Born’s rule for quantum measurement}

\author{Lotte Mertens}
\affiliation{Institute for Theoretical Physics Amsterdam,
University of Amsterdam, Science Park 904, 1098 XH Amsterdam, The Netherlands}
\affiliation{Institute for Theoretical Solid State Physics, IFW Dresden, Helmholtzstr. 20, 01069 Dresden, Germany}
\author{Jasper van Wezel}
\affiliation{Institute for Theoretical Physics Amsterdam,
University of Amsterdam, Science Park 904, 1098 XH Amsterdam, The Netherlands}

\date{\today}

\begin{abstract}
The argument of environment-assisted invariance (known as envariance) implying Born's rule is widely used in models for quantum measurement to reason that they must yield the correct statistics, specifically for linear models. However, it has recently been shown that linear collapse models can never give rise to Born's rule. Here, we address this apparent contradiction and point out an inconsistency in the assumptions underlying the arguments based on envariance. We use a construction in which the role of the measurement machine is made explicit and show that the presence of envariance does \emph{not} imply every measurement will behave according to Born's rule. Rather, it implies that every quantum state allows a measurement machine to be constructed, which yields Born's rule when measuring that particular state. This resolves the paradox and is in agreement with the recent result of objective collapse models necessarily being non-linear. 
\end{abstract}

\maketitle

\section{Introduction}
It has recently been shown that linear models for objective collapse (i.e. models whose time evolution is generated by a linear operator) cannot give rise to Born's rule ~\cite{no-linear}. This is surprising, given that they fall into a class of models in which the emergence of Born's rule was previously suggested to be unavoidable~\cite{Zurek,vanwezelprb}. The suggested proof for the emergence of Born's rule was first formulated in the context of decoherence and is based on the possibility of quantum states entangling with an external environment~\cite{Zurek}. However, the assumptions entering the suggested proof were claimed not to depend on the actual presence, influence, or dynamics of any environmental states. Essentially the same reasoning has therefore also been applied in several other well-known approaches to the quantum measurement problem, including the pilot wave and many-worlds theories~\cite{Deutsch_99,Wallace_03,Wallace_12,Valentini,Saunders,vanwezelprb}.

In fact, the arguments based on envariance~\cite{Zurek} can be applied to any model for quantum state reduction, regardless of whether it involves an explicit environment and regardless of whether it is linear or non-linear. The envariance-based result, therefore, appears to contradict the recent proof that general two-state time evolution can only yield the emergence of Born's rule if the dynamics is non-linear~\cite{no-linear}. Here, we resolve the paradox by revealing an assumption in the envariance-based analysis, which significantly limits its applicability.

\section{Envariance}
In order to contrast the arguments for the emergence of Born's rule from envariance with the observation that Born's rule cannot arise in linear models for quantum state reduction, we will first give a brief sketch of the suggested envariance-based proof. We will use the concept of envariance as introduced by Zurek~\cite{Zurek}. That is, if an operation $\hat{U}_{\text{S}}$ acting on a local system S can be undone by an operation $\hat{U}_{\text{E}}$ acting solely on an external environment E, such that the combined system SE is unchanged under $\hat{U}=\hat{U}_{\text{E}}\hat{U}_{\text{S}}$, the state is said to be envariant under the operation $\hat{U}_{\text{S}}$~\cite{Zurek}.

The principal idea is then that the statistical properties of any local measurement on a quantum state cannot be influenced by physical operations taking place at a causally disconnected location or event. If this condition were violated, instantaneous, faster-than-light communication would be possible. Applying the causality condition to quantum states envariant under various types of operations imposes fundamental constraints on statistics of measurement outcomes.

In the following, we will consider quantum states for the local system written as $\ket{\Psi_S} = \sum_{i=1}^N c_i \ket{S_i}$, with complex components $c_i$ for the orthonormal basis states $\{ \ket{S_i} \}$. Entangling the local system with environmental degrees of freedom results in a state of the form $\ket{\Psi_{SE}}= \sum_{i=1}^N c_i \ket{S_i}\ket{E_i}$. Unless otherwise specified, any measurements we consider will be on the local system without involving the environment (which might be in a causally disconnected location). 

Starting from a superposition over two system states, the phase of the complex components in the quantum state can be shown not to influence the statistics of measurement outcomes, due to envariance. Consider the following entangled state:
\begin{align}
    & \ket{\psi} = \alpha \ket{S_0, E_a} + \beta \ket{S_1, E_b}. 
    \label{eq:alphabeta}
\end{align}
Here, $\ket{S_0}$ and $\ket{S_1}$ are basis states for the local quantum system. The kets $\ket{E_a}$ and $\ket{E_b}$ represent states of an environmental degree of freedom that, once entangled with the system, is taken to a far-away location where it is causally disconnected from the system.

The state $\ket{\psi}$ is envariant under the phase rotation $\hat{U}_S \ket{S_0} = e^{i\phi} \ket{S_0}$, since the effect of this operation on the state $\ket{\psi}$ can be undone by the environmental operation $\hat{U}_E \ket{E_a} = e^{-i\phi} \ket{E_a}$. That is, $\hat{U}_E\hat{U}_S\ket{\psi}=\ket{\psi}$. This directly implies that the phase of the complex coefficient $\alpha$ cannot influence the statistics of measurements on any local property of the state, because that phase can be altered by an actor operating solely on the causally disconnected environment.

A similar argument also holds for the phase of $\beta$, and we thus arrive at the result that the statistical outcomes of quantum measurement can depend only on the absolute values of its wave function components. This result places a constraint on the properties of any model for quantum measurement: for a state of the form $\sum_{i=1}^N c_i \ket{S_i}\ket{E_i}$, the statistics of a local measurement on S can only depend on the absolute values $\abs{c_i}$. If this restriction is violated, faster-than-light communication between causally disconnected events would be possible.

\section{Quantum measurement}
By itself, envariance is a mathematical property of particular entangled quantum states. The envariance of states occurring in particular models of quantum measurement may be used to argue that the possible outcomes of these models are constrained. For example, the arguments of the previous section could be used to argue that in particular models for quantum measurement, the probability of finding any outcome does not depend on the phases in the initial state decomposition. Whether or not arguments based on envariance can be applied to any specific measurement model, depends on the detailed properties of that model.

In Ref.~\onlinecite{Zurek}, for example, it is shown using arguments of envariance, that the existential interpretation of quantum measurement satisfies the particular causality condition of probabilities being independent of the phases of wave function components. Essentially the same arguments have been used to argue more generally that pilot wave and many worlds interpretations (in combination with decoherence) also satisfy it~\cite{Deutsch_99,Wallace_03,Wallace_12,Valentini,Saunders}, as do objective collapse theories based on spontaneous symmetry breaking~\cite{vanwezelprb}. The essential ingredients in all of these demonstrations were identified in Ref.~\onlinecite{vanwezelprb} to be:
\begin{enumerate}
    \item The measurement process can, at least in principle, be divided into pre-measurement and registration.
    \item There is a preferred basis in which registration takes place.
    \item The selection of a particular preferred basis state as the measurement outcome is probabilistic.
    \item The registration is a local process that is not influenced by events causally disconnected from the measurement machine.
    \item The probability of selecting any particular measurement outcome does not depend on any physical property of either the system being measured or the measurement machine used to measure it.
    \item The measurement process does not \emph{a priori} favour any particular outcome.
\end{enumerate}

In fact, these six conditions turn out to allow the application of envariance-based arguments in general. That is, if a model for quantum measurement obeys these conditions, we can employ envariance to constrain the probability distribution of measurement outcomes.

To see how this works for the particular constraint that measurement outcomes do not depend on the phases of wave function components, consider a pair of entangled states:
\begin{align}
    & \ket{\psi} = \alpha \ket{S_0, P_0, E_a} + \beta \ket{S_1, P_1, E_b}, \notag \\
    & \ket{\chi} = \alpha \ket{S_0, P_0, E_c} + \beta \ket{S_1, P_1, E_d}.
    \label{eq:alphabeta2}
\end{align}
Here, $\ket{S_0}$ and $\ket{S_1}$ are the basis states of the quantum system being measured. The states $\ket{P_0}$ and $\ket{P_1}$ are (pointer) states of the measurement machine that became entangled with the system during pre-measurement (condition 1)~\cite{sneed1966neumann}. Finally, the kets $\ket{E_a}$, $\ket{E_b}$, $\ket{E_c}$, and $\ket{E_d}$ are states of an environment degree of freedom that may be located far from the system and measurement machine.

The registration phase of the measurement process will select one of the pointer states as the measurement outcome (condition 2) in a probabilistic fashion (condition 3). It does not matter whether this happens through a dynamical evolution of the state itself, as in objective collapse theories~\cite{vanwezelprb}, or by assigning an observer or register to one of the branches, as in relative-state-based interpretations~\cite{Zurek,Deutsch_99,Wallace_03,Wallace_12,Valentini,Saunders}. 

Conditions 4, 5, and 6 ensure that the probability for any particular measurement outcome to be registered cannot depend on what the kets appearing in Eq.~\eqref{eq:alphabeta2} represent physically. The probabilities must therefore be determined entirely by the coefficients $\alpha$ and $\beta$. This implies in particular that the probability for $\ket{S_0,P_0,E_a}$ being registered starting from the state $\ket{\psi}$ must equal that of $\ket{S_0,P_0,E_c}$ being registered starting from $\ket{\chi}$. Finally, condition 4 ensures that this is true even for the special case in which $\ket{E_c} \to e^{i\varphi}\ket{E_a}$ and $\ket{E_d} \to \ket{E_b}$, showing that the probability for registering $\ket{S_0,P_0,E_a}$ cannot depend on the phase of the $\alpha$ coefficient.

\section{Swaps}
Following the central steps in the suggested derivation of Born's rule in ref.~\onlinecite{Zurek}, we continue by considering the entangled state of equation~\eqref{eq:alphabeta}, but in the special case with equal values for the coefficients:
\begin{align}
    & \ket{\psi} = \alpha \left( \ket{S_0, E_a} + \ket{S_1, E_b} \right).
    \label{eq:alphaonly}
\end{align}
Besides the envarinace under local phase rotations, this state is also envariant under the unitary swap operation defined by $\hat{U}_{\text{S}} = \ket{S_0}\bra{S_1} + \ket{S_1}\bra{S_0}$. This swap of the system state can be undone by a unitary swap on the environment state defined by $\hat{U}_{\text{E}} = \ket{E_a}\bra{E_b} + \ket{E_b}\bra{E_a}$, so that $\hat{U}_{\text{E}} \hat{U}_{\text{S}} \ket{\psi} = \ket{\psi}$, which demonstrates the envariance.

The swap operation on environmental states that are causally disconnected from the system cannot influence the statistics of local measurements on the system. This means that the statistics obtained when measuring $\ket{\psi}$ must equal those obtained when measuring $\hat{U}_{\text{E}}\ket{\psi}$. But because in this particular case with equal coefficients $\hat{U}_{\text{E}}\ket{\psi}=\hat{U}_{\text{S}}\ket{\psi}$, the probability for obtaining a measurement outcome associated with $\ket{S_0}$ must equal the probability for finding the $\ket{S_1}$ outcome. 

In other words, any model that does not yield equal probabilities for measurement outcomes starting from an equal-weight superposition allows faster-than-light communication. Again, it can be argued that imposing the conditions 1-6 of the previous section on models for quantum measurement suffices to ensure this condition will be adhered to. This can be seen by considering the pair of states:
\begin{align}
    & \ket{\psi} = \alpha \ket{S_0, P_0, E_a} + \beta \ket{S_1, P_1, E_b}, \notag \\
    & \ket{\chi} = \beta \ket{S_0, P_0, E_a} + \alpha \ket{S_1, P_1, E_b}.
    \label{eq:alphabeta3}
\end{align}
This state can be considered the result of pre-measurement (condition 1), while conditions 2 and 3 will ensure that registration results in the probabilistic selection of one of the pointer states. Conditions 4, 5, and 6 again ensure that the probability for any particular measurement outcome is determined entirely by the $\alpha$ and $\beta$ coefficients. 

For Eq.~\eqref{eq:alphabeta3}, this implies in particular that the probability for $\ket{S_0,P_0,E_a}$ being registered starting from the state $\ket{\psi}$ must equal that of $\ket{S_1,P_1,E_b}$ being registered starting from $\ket{\chi}$. Condition 4 ensures that this is true even for the special case in which $\beta \to \alpha$. But since, in that case, the two states become equal, they must additionally have the same probability of registering $\ket{S_0, P_0, E_a}$. We thus find that any model for measurement meeting conditions 1-6 will find equal probabilities for registering any particular final state starting from an equal-weight superposition.

Extending the argument, we can consider a state involving arbitrarily many components:
\begin{align} 
\label{eq: initial}
    \ket{\psi} = \sum_{k=1}^N \alpha_k \ket{S_k, E_k}.
\end{align}
If the weights $\alpha_k$ are equal for any pair of labels $k'$ and $k''$, then the state $\ket{\psi}$ is left invariant by the consecutive swaps $\hat{U}_{\text{S}} = \ket{S_{k'}}\bra{S_{k''}} + \ket{S_{k''}}\bra{S_{k'}}$ and $\hat{U}_{\text{E}} = \ket{E_{k'}}\bra{E_{k''}} + \ket{E_{k''}}\bra{E_{k'}}$. Using similar arguments as above, we then conclude that any subset of states with equal weights within a larger superposition must all have equal probabilities of being registered in a local measurement on the system. 

Notice that none of the arguments in this or the previous section require the environmental states to actually exist or be present~\cite{Zurek}. In fact, since local actions on the environment cannot influence the statistics of local measurement outcomes on the system, we could consider an extreme case in which the environmental degree of freedom is destroyed (without measuring it) before the system is measured. Since the destruction of the environment cannot influence the statistics observed in the system, the probability of registering any particular outcome must be independent of whether the environment exists.

\section{Born's rule?}
Finally, the suggested argument for Born's rule being implied by envariance starts from a superposition with unequal weights:
\begin{align}
    \ket{\psi} = \alpha \ket{S_0,E_a} + \beta \ket{S_1,E_b}.
    \label{eq:general}
\end{align}
We may assume the environmental Hilbert space to be arbitrarily large. It is then always possible to identify an alternative basis $\{\ket{E'}\}$ for the environmental states, in which the full state can be expressed as an equal-weight superposition:
\begin{align}
    \ket{\psi} = \sqrt{\frac{1}{N}} \left[ \sum_{i=1}^n \ket{S_0,E'_i} + \sum_{j=n+1}^{n+m} \ket{S_1,E'_j} \right].
    \label{eq:zurekstateapp}
\end{align}
Here, the rational fractions $n/N$ and $m/N$ can be made to approximate the real numbers $\alpha^2$ and $\beta^2$ with arbitrary precision~\cite{Zurek}. Because the weights of all components in this state are equal, causality guarantees equal probabilities for registering any one of them. 

To see this explicitly, we need to identify two swap operations whose product leaves the state invariant. In this case, however, the swap of system states, $\hat{U}_{\text{S}} = \ket{S_{0}}\bra{S_{1}} + \ket{S_{1}}\bra{S_{0}}$, cannot be undone by a swap operation on the environment $E'$. The only exception is the special case $m=n$, which would imply we had an equal-weight superposition with $\alpha=\beta$, to begin with. To find a combination of operations that does leave the state invariant, we need to consider the possible existence of a second environment $E''$, which we may assume to be causally disconnected from both the system and the first environment:
\begin{align}
    \ket{\psi} = & \sqrt{\frac{1}{N}} \left[ \sum_{i=1}^n \ket{S_0,E'_i,E''_i} + \sum_{j=n+1}^{n+m} \ket{S_1,E'_j,E''_j} \right].
    \label{eq:zurekstateapp2}
\end{align}
In this state, a combined swap on the system and the first environment, $\hat{U}_{\text{S}} = \ket{S_{0}, E'_i}\bra{E'_{j},S_1} + \ket{S_{1}, E'_j}\bra{E'_i,S_{0}}$, can be undone by a local swap on the second environment, $\hat{U}_{\text{E}} = \ket{{E''_i}}\bra{{E''_j}} + \ket{{E''_j}}\bra{{E''_i}}$. Thus the probabilities of registering any of the states $\ket{S_0,E'_i,E''_i}$ or $\ket{S_1,E'_j,E''_j}$ must all be equal.

The final step in the suggested proof is then to argue that because all states $\ket{S_0,E'_i,E''_i}$ contain the system state $\ket{S_0}$, and because all of them are orthogonal, the probability of registering a measurement outcome associated with $\ket{S_0}$ is equal to $n$ times the probability for registering any one of the states $\ket{S_0,E'_i,E''_i}$. That is, the probability of finding the measurement outcome associated with $\ket{S_0}$ is suggested in Ref.~\onlinecite{Zurek} to equal $n/N$, in accordance with Born's rule.

Notice that if the reasoning in the previous paragraph holds, it would mean that causality (in the form of envariance) implies Born's rule \emph{and} that any model for measurement meeting conditions 1-6 would be guaranteed to give rise to Born's rule. However, it was shown by explicit construction in Ref.~\onlinecite{no-linear} that there are objective collapse models meeting conditions 1-6 which nonetheless do not yield measurement statistics agreeing with Born's rule. In fact, it was shown that linear objective collapse models could not possibly produce Born's rule, even though conditions 1-6 do not exclude linear theories. This seems to leave us with a contradiction. 

\section{Measurement machines}
We will resolve the paradox by explicitly including the measurement machine and considering the states after measurement in the final steps of the suggested proof. First, focus on a particular instance of Eq.~\eqref{eq:general}, in which the state after pre-measurement but before registration of the measurement outcome is given by $\ket{\psi_1} = \sqrt{3}/2 \ket{S_0,P_0,E_0} + 1/2 \ket{S_1,P_1,E_1}$. As before, we can argue that there is always an alternative basis for the environmental Hilbert space in which this state can be written as:
\begin{align}
    \ket{\psi_1} &= 1/2 \left( \ket{S_0, P_0,E'_a} + \ket{S_0, P_0, E'_b} + \right. \notag \\
    & ~~~~~~~~~~ + \left.
    \ket{S_0, P_0,E'_c} + \ket{S_1, P_1, E'_d} \right).
    \label{eq:psi1}
\end{align}
The components in this state now have equal weights. 

Next, consider the registration phase and the eventual outcome of the measurement process. First, assume the measurement process acting on the machine with pointer states $\ket{P_0}$ and $\ket{P_1}$ is purely local and does not affect the environment. If the measurement machine then ends up registering the measurement outcome $0$, the state corresponding to the outcome will be given by the projection~\cite{sneed1966neumann}:
\begin{align}
    &\mathcal{I}_{S}\otimes\ket{P_0}\bra{P_0}\otimes\mathcal{I}_{E'} \, \ket{\psi_1} \notag \\ 
    &= 1/2 \left( \ket{S_0, P_0,E'_a} + \ket{S_0, P_0, E'_b} +
    \ket{S_0, P_0,E'_c} \right).
    \label{eq:out1}
\end{align}
Here, $\mathcal{I}_S$ is the identity operator on the system states, and $\mathcal{I}_{E'}=\mathcal{I}_E$ is the identity operator for the environment. We do not make any assumptions about the interpretations of quantum mechanics. If the measurement were described by an objective collapse theory, the (normalised) state of Eq.~\eqref{eq:out1} would be the actual state after measurement. If instead, we take a relative state interpretation, Eq.~\eqref{eq:out1} represents one component of the final state superposition, which is then entangled --\emph{as a whole}-- with a register or memory state storing the outcome $0$ for the measurement.

Notice that the measured state of Eq.~\eqref{eq:out1} is not one of the components of the equal-weight superposition of Eq.~\eqref{eq:psi1}. Rather, it corresponds precisely to $\ket{S_0,P_0,E_0}$ which appeared in $\ket{\psi_1}$ with weight $\sqrt{3}/2$. Because the selected final state is not one of the equal weight components, we cannot use envariance to say anything about the likeliness with which it is selected.

To invoke envariance, we need to consider a measurement process resulting in a measurement outcome that corresponds to just one of the components $\ket{S_0, P_0,E'_a}$, $\ket{S_0, P_0,E'_b}$, or $\ket{S_0, P_0,E'_c}$. This requires the state after registration (obtained either objectively or relative to a memory state) to be given by a projection operator on the environment as well as the pointer state\footnote{Notice that projecting on the environment alone would suffice. Since a measurement machine should have a pointer pointing out the measurement result, however, we include a projection on the pointer states.}:
\begin{align}
    \left[ \mathcal{I}_{S}\otimes\ket{P_0}\bra{P_0}\otimes\ket{E'_j}\bra{E'_j} \right] \ket{\psi_1} = \frac{1}{2} \ket{S_0, P_0,E'_j}.
    \label{eq:out2}
\end{align}
For the states registered by a measurement machine enacting these types of projections, the arguments of the previous section can be applied. We then find equal probabilities for obtaining any of the four components in Eq.~\eqref{eq:psi1}. Notice, however, that this projection operation affects both the pointer \emph{and} the environment. The corresponding measurement machine thus violates condition 4 by acting non-locally. Moreover, the environmental states $\ket{E'}$ cannot be argued to be argued to be hypothetical or included for the sake of argument only. For envariance-based arguments to be applied to the measurement process, it necessarily needs to projects onto precisely the states $\ket{E'}$ that create an equal-weight superposition.

It might be argued that the environmental states $\ket{E'}$ could actually exist and that, moreover, they might actually be local to the pointer, by somehow arguing that the measurement machine includes the environmental states $\ket{E'}$. Even in that case, however, a problem remains. To see this, consider an alternative initial state, which is given after pre-measurement by $\ket{\psi_2} = \sqrt{2/3} \ket{S_0,P_0,E_0} + \sqrt{1/3} \ket{S_1,P_1,E_1}$. Notice that the environmental states $\ket{E_0}$ and $\ket{E_1}$ here are the exact same states that appeared in our earlier discussion of $\ket{\psi_1}$. This is necessary because the measurement machine ought to entangle $\ket{S_0,P_0}$ with the same environmental state $\ket{E_0}$ regardless of the weight with which $\ket{S_0,P_0}$ appears in the initial wave function. If this were not the case, the Hamiltonian describing the pointer-environment interaction would not be a linear operator.

In the alternative basis of environmental states $\ket{E'}$ that we introduced to write $\ket{\psi_1}$ as an equal weight superposition, $\ket{\psi_2}$ becomes:
\begin{align}
    \ket{\psi_2} &= \sqrt{2}/3 \left( \ket{S_0, P_0,E'_a} + \ket{S_0, P_0, E'_b} + \right. \notag \\
    & ~~~~~~~~~~ + \left.
    \ket{S_0, P_0,E'_c} \right) + \sqrt{1/3} \ket{S_1, P_1, E'_d}.
    \label{eq:psi2a}
\end{align}
Clearly, this is not an equal weight superposition and we can not use envariance to say anything about the likeliness of measurement selecting any particular component in this state. 

In order to invoke envariance, we would first need to define yet a different basis for the environment, in which the state can be written in the form:
\begin{align}
    \ket{\psi_2} &= \sqrt{1/3} \left( \ket{S_0, P_0,E''_A} + \ket{S_0, P_0, E''_B} \right. \notag \\
    &~~~~~~~~~~ + \left. \ket{S_1, P_1, E''_C} \right).
    \label{eq:psi2b}
\end{align}
The measurement that is guaranteed by envariance to have equal probabilities of registering any outcome is one that corresponds to the projection operators $\mathcal{I}_{S}\otimes\ket{P_0}\bra{P_0}\otimes\ket{E''_J}\bra{E''_J}$. This is a different projector than the one in Eq.~\eqref{eq:out2}, and thus it describes a different physical measurement machine being used. In other words, there is no single measurement process (projecting into one particular basis) that allows for the envariance-based arguments to be applied regardless of the initial wave function.

In fact, each initial state configuration of the same physical system $\ket{S}$ requires a different measurement machine in order for envariance to guarantee that its measurement outcomes adhere to Born's rule. Moreover, to know which measurement machine will yield the correct statistics, we need to know the exact weights of all wave function components in the initial state. Only then can we identify the basis of environmental states that yields an equal-weight superposition, which we need to identify the measurement machine projecting onto that basis. Clearly, this does not correspond to physical experience, in which the same measurement machine can be used to measure any state of a system.

\section{Conclusion}
In conclusion, we showed that the arguments based on envariance that were previously argued to imply Born's rule~\cite{Zurek}, in fact, only show that for each initial system state, it is possible to define a (non-local) measurement machine that projects onto a combined system-environment state with Born rule (equal) probabilities. This definition of the measurement machine is different for each initial system state to be measured.

We also demonstrated that the envariance-based arguments cannot be used to predict the probabilities for measurement outcomes starting from general initial states using a local measurement machine projecting onto system states only. Envariance, therefore, does not generally imply Born's rule. This agrees with the recent observation that linear objective collapse models cannot give rise to Born's rule~\cite{no-linear}, even though they obey conditions 1-6 defined above. Because envariance cannot be used to derive Born's rule, the conditions do not imply Born's rule, and the apparent contradiction is resolved.

\end{document}